\documentstyle[preprint,aps,epsfig]{revtex}
\renewcommand{\today}{14 February, 1996}

\newcommand{\nc}{\newcommand}
\nc{\be}{\begin{equation}}
\nc{\ee}{\end{equation}}
\nc{\bea}{\begin{eqnarray}}
\nc{\eea}{\end{eqnarray}}
\nc{\beas}{\begin{eqnarray*}}
\nc{\eeas}{\end{eqnarray*}}
\nc{\noi}{\noindent}
\nc{\sD}{\not \! \! D}
\nc{\s}[1]{\not \! #1}
\nc{\non}{\nonumber}
\nc{\bb}{\bibitem}
\nc{\rw}{$\rho\!-\!\omega$ }
\nc{\oeps}{${\cal O}(\epsilon )$}
\nc{\lf}{\left}
\nc{\r}{\right}
\nc{\mb}[1]{\makebox[#1]{}}
\nc{\pa}{\partial}
\nc{\sA}{\not \! \! A}
\nc{\newsec}[1]{\section{#1}\mb{0.5cm}}
\nc{\h}{\frac{1}{2}}
\nc{\ra}{\rightarrow}
\nc{\la}{\leftarrow}
\nc{\ep}{$e^+e^-\ra\pi^+\pi^-\;$}
\def\mathunderaccent#1{\let\theaccent#1\mathpalette\putaccentunder}
\def\putaccentunder#1#2{\oalign{$#1#2$\crcr\hidewidth
\vbox to.2ex{\hbox{$#1\theaccent{}$}\vss}\hidewidth}}

\nc{\ti}{\mathunderaccent\tilde}
\nc{\M}{{\cal M}}

\begin{document}
\thispagestyle{empty}
\begin{flushright}
Published in Phys. Lett. B376 (1996) 19. \\
ADP-95-50/T197 \\
hep-ph/9601309
\end{flushright}

\begin{center}
{\large{\bf Analysis of rho-omega 
interference in the pion form-factor}} \\
\vskip48pt
Kim Maltman$^{a,b}$, H.B.~O'Connell$^{b}$ and  A.G.~Williams$^{b,c}$ \\
\vskip24pt
{\it $^{a}$Department of Mathematics and Statistics, York University,
4700 Keele St., \\ North York, Ontario, Canada M3J 1P3 }\\
{\it $^{b}$Department of Physics and Mathematical Physics, 
University of Adelaide 5005, Australia}\\
{\it $^{c}$Institute for Theoretical Physics,
University of Adelaide 5005, Australia } \\
\vskip24pt
\today

\begin{abstract}
The formalism underlying the analysis of $e^+e^-\rightarrow\pi^+\pi^-$ in the
$\rho-\omega$ interference region is carefully revisited.  
We show that the standard neglect of the pure $I=0$ omega, 
$\omega_I$, ``direct'' coupling to $\pi\pi$ is not valid, and
extract those combinations of the direct coupling and 
$\rho$-$\omega$ mixing allowed by experiment.  The latter is shown
to be only very weakly constrained by experiment, and we conclude 
that data from the $e^+e^- \rightarrow \pi^+\pi^-$ interference
region {\em cannot} be used to fix the value 
of \rw mixing 
in a model-independent way unless the errors on the experimental phase can be
significantly reduced.  Certain other modifications of the usual
formalism necessitated by the unavoidable momentum-dependence
of \rw mixing are also discussed.
\end{abstract} 
 \end{center}
\vfill
\begin{flushleft}
E-mail: {\it hoconnel, awilliam@physics.adelaide.edu.au; 
FS300175@sol.yorku.ca} \\
Keywords: vector mesons, pion form-factor, mixing, isospin.\\
PACS: 11.20.Fm, 12.40.Vv, 13.60.Le, 13.65.+i

\end{flushleft}

\newpage


The cross-section for \ep{} in the \rw resonance region displays a narrow
interference shoulder resulting from the superposition of narrow resonant
$\omega$ and broad resonant $\rho$ exchange amplitudes \cite{Barkov}.  The
strength of the $\omega$ ``interference'' amplitude has generally been taken to
provide a measurement of $\rho_I$-$\omega_I$ mixing (where $\rho_I$, $\omega_I$
are the pure isovector $\rho$ and isoscalar $\omega$ states) \cite{CB,OPTW2}. 
The extracted mixing has then been used to generate $\rho_I$-$\omega_I$ mixing
contributions to various few-body observables \cite{HM,MNS,nuclear}, a program
which, combined with estimates for other sources of isospin-breaking, produces
predictions for few-body isospin breaking in satisfactory accord with
experiment \cite{MNS}. The phenomenological success, for those observables for
which $\rho_I$-$\omega_I$ contributions are significant, rests, inextricably,
on two assumptions, (1) that the interference amplitude is dominated by
$\rho_I$-$\omega_I$ mixing (i.e., negligible ``direct''
$\omega_I\rightarrow\pi\pi$ contribution to the physical $\omega$ decay
amplitude) and (2) that the resulting mixing amplitude is independent of
momentum-squared, so the extracted value can be used unchanged in
meson-exchange forces in few-body systems, where $q^2<0$.

The neglect of ``direct'' $\omega_I\rightarrow\pi\pi$ coupling (i.e.,
coupling which does not go via mixing with the $\rho_I$) can actually be
re-interpreted physically, this re-interpretation simultaneously providing the
conventional justification for taking the $\rho_I$-$\omega_I$ self-energy,
$\Pi^{\rho\omega}$, to be real in modern analyses of \ep \cite{Renard,review}. 
As will become clear below, however, corrections to the underlying
argument, usually thought to be small, have unexpectedly large effects
on the extraction of the \rw mixing contribution from experimental data.

The assumption of the $q^2$-independence of $\Pi^{\rho\omega}(q^2)$
is more problematic \cite{CS,OPTW}.  
In general, one knows that a
system of, e.g., nucleons, vector mesons and pseudoscalar mesons, 
can be described by an effective low-energy Lagrangian,
constructed so as to be compatible with QCD
(e.g., one might think of the effective chiral
Lagrangian, ${\cal L}_{\rm eff}$, obtainable via the Coleman-Callan-Wess-Zumino
construction \cite{CCWZ}).  Such a Lagrangian, involving terms of arbitrarily
high order in derivatives, will produce momentum-dependence in all observables
which can in principle become momentum-dependent.
This has been seen explicitly for the
off-diagonal (mixing) elements of meson propagators by a number of authors,
employing various models \cite{models,MTRC}, as well as QCD sum rule and
Chiral Perturbation Theory (ChPT) techniques \cite{tech}.  
Such $q^2$-dependence has also been shown to be 
consistent with the usual vector meson
dominance (VMD) framework \cite{OWBK}.  The possibility \cite{CM} that an
alternative choice of interpolating fields might, nonetheless, correspond to
the standard assumption of $q^2$-independence has been shown to be incompatible
with the constraints of unitarity and analyticity \cite{KM1}.  It is thus
appropriate to revisit and generalize the usual analysis.

As has been known for some time, to obtain properties of unstable
particles which are process-independent and physically meaningful, one 
determines the locations of the resonance poles in the amplitude under
consideration, and makes expansions about these pole locations \cite{pole}.
The (complex) pole locations are properties of the S-matrix and hence {\em
independent of the choice of interpolating fields}, and the separate terms in
the Laurent expansion about the pole position have well-defined physical
meaning \cite{pole}.  The importance of such an ``S-matrix'' formalism for
characterizing resonance properties has been stressed recently by a number of
authors in the context of providing gauge- and process-independent definitions
of the $Z^0$ mass and width in the Standard Model \cite{Sirlin,polepapers}. For
our purposes this means that: (1) the ``physical'' \{$\rho$, $\omega$\} fields
are to be identified as those combinations of the \{$\rho_I$, $\omega_I$\}
fields containing the corresponding S-matrix poles and (2) to analyze \ep one
should include both resonant terms involving the complex $\rho$ and $\omega$
pole locations (and hence constant widths) and ``background'' (i.e.
non-resonant) terms.  In quoting experimental results we will, therefore,
restrict ourselves to analyses which, as closely as possible, satisfy these
requirements.  To our knowledge, only one such exists: the fifth fit of
Ref.\cite{Bernicha} (performed explicitly in the S-matrix formalism, though
without an $s$-dependence to the background).  As stressed in
Ref.~\cite{Bernicha}, using the S-matrix formalism, one finds a
somewhat lower real part for the (complex) $\rho$ pole position 
($\hat{m}_\rho =757.00 \pm 0.59,\; \Gamma_\rho =143.41 \pm 1.27$ MeV) than is
obtained in conventional, non-S-matrix formalism treatments.  For comparison
below we will also employ the results of the second fit of the more
conventional (but non-S-matrix) formalism of Ref.  \cite{Benayoun}, which
employs an $s$-dependent background, an $s$-dependent 
$\rho$ width, 
and imposes the (likely too large)
Particle Data Group value for the
$\rho$ mass by hand. 

Let us turn to the question of \rw mixing in the presence of a $q^2$-dependent
off-diagonal element of the self-energy matrix.  We shall work consistently to
first order in isospin breaking (generically, ${\cal O}(\epsilon )$), 
which will mean to first order in
$\Pi_{\rho\omega}$. The dressing of the bare, two-channel meson propagator has
been treated in Ref.~\cite{OPTW}.  

As we consider vector mesons coupled to conserved currents, we can 
replace $ D_{\mu\nu}(q^2)$ by
$ -g_{\mu\nu}D(q^2)$.
We refer to $D(q^2)$ as the ``scalar propagator''.  
We assume that the
isospin-pure fields $\rho_I$ and $\omega_I$ have already been 
renormalized, i.e.,
that the relevant counterterms have been absorbed into the mass and
wavefunction renormalizations.  Taking then the full expression for the dressed
propagator and keeping terms
to \oeps , one finds
\be 
D^I(q^2)=\left(\begin{array}{cc}
D^I_{\rho\rho} & D^I_{\rho\omega} \\ 
D^I_{\rho\omega} & D^I_{\omega\omega} \end{array}\right)=
\left(\begin{array}{cc}
(q^2-\Pi_{\rho\rho}(q^2))^{-1} & D^I_{\rho\omega}(q^2) \\ 
D^I_{\rho\omega}(q^2) &
(q^2-\Pi_{\omega\omega}(q^2))^{-1} \end{array}\right),  
\label{two} 
\ee 
where the renormalized self-energies $\Pi_{kk}(q^2)\rightarrow
m^2_k$ as $q^2\rightarrow m^2_k$.  Defining $\Pi_{kk}^{(0)}(q^2)
=\Pi_{kk}(q^2)-m^2_k$, we then have $\Pi_{kk}^{(0)}(q^2)=
{\cal O}[(q^2-m^2_k)^2]$.
{From} the complex pole positions, $m_k^2$, we define the (real)
mass ($\hat{m}_k$) and width ($\Gamma_k$) via,
$m_k^2\equiv\hat{m}_k^2-i\, \hat{m}_k\Gamma_k.$
To \oeps ,  
$D_{\rho\omega}^I(q^2)$, is then \cite{OPTW}
\be
D_{\rho\omega}^I(q^2)={\Pi_{\rho\omega}(q^2)\over
(q^2-m_\rho^2-\Pi_{\rho\rho}^{(0)}(q^2))(
q^2-m_\omega^2-\Pi_{\omega\omega}^{(0)}(q^2))}
=D^I_{\rho\rho}(q^2)\Pi_{\rho\omega}(q^2)D^I_{\omega\omega}(q^2),
\label{three}
\ee which
contains both a broad $\rho$ resonance and narrow $\omega$ resonance piece. 

As explained above, the physical $\rho$ and $\omega$ fields are 
defined to be
those combinations of the $\rho_I$ and $\omega_I$ for which
only the diagonal elements of
the propagator matrix contain poles,
in the ${\rho ,\omega}$ basis.
This definition 
is, in fact, 
implicit in the standard interpretation of the \ep experiment,
which associates the broad resonant part of the full amplitude with the $\rho$
and the narrow resonant part with the $\omega$.  
Using different linear combinations of $\rho_I$, $\omega_I$, 
(call them $\rho^\prime$, 
$\omega^\prime$) than those given above ($\rho$, $\omega$),
one would find also narrow resonant structure in the
off-diagonal element of the vector meson propagator in the \{$\rho^\prime$,
$\omega^\prime$\} basis, preventing,
for example, the association of the narrow resonant
behaviour with the $\omega^\prime$ pole term alone.

We define the transformation between the physical and isospin pure bases by 
(to ${\cal O}(\epsilon)$)
\be
\rho=\rho_I -\epsilon_1\,\omega_I,\hskip0.5cm
\omega=\omega_I+\epsilon_2\,\rho_I\label{four}
\ee
where, in general, $\epsilon_1\not= \epsilon_2$ when the
mixing is $q^2$-dependent.  With $D_{\rho \omega}^{\mu
\nu}(x-y)\equiv -i\langle0|T(\rho^\mu(x)\omega^\nu(y))\vert 0\rangle$,
one then has for the scalar propagator, to \oeps ,
\begin{equation}
D_{\rho\omega}(q^2)=D^I_{\rho\omega}(q^2)
-\epsilon_1D^I_{\omega\omega}(q^2)+\epsilon_2D^I_{\rho\rho}(q^2).
\label{five}
\end{equation}
The condition that
$D_{\rho\omega}(q^2)$
contain no $\rho$ or $\omega$ pole then 
fixes $\epsilon_{1,2}$ to be
\be
\epsilon_1={\Pi_{\rho\omega}(m_\omega^2)\over m_\omega^2-
m_\rho^2-\Pi_{\rho\rho}^{(0)}(m_\omega^2)},\hskip0.5cm
\epsilon_2={\Pi_{\rho\omega}(m_\rho^2)\over m_\omega^2-
m_\rho^2+\Pi_{\omega\omega}^{(0)}(m_\rho^2)}\ .\label{six}
\ee

When $\Pi^{\rho\omega}(q^2)$ is $q^2$-dependent, we 
thus see explicitly that
$\epsilon_1\not=\epsilon_2$; the relation between the isospin-pure and physical
bases is not a simple rotation.  This is a universal feature of $q^2$-dependent
mixing in field theory.  Recall that $\Pi_{\rho \rho}^{(0)}(q^2)$ 
and $\Pi_{\omega\omega}^{(0)}(q^2)$ 
vanish by definition as $q^2\ra m^2_{\rho ,\omega}$ 
at least as fast as $(q^2- m_{\rho ,\omega}^2)^2$.
The usual assumption is that these two quantities are zero in the vicinity of
the resonance region, which leads to the standard Breit-Wigner
form for the vector meson propagators.  $\Pi_{\rho
\rho}^{(0)} (q^2)$ and $\Pi_{\omega \omega}^{(0)} (q^2)$ 
are, of course, 
momentum-dependent in general
since the vector propagators must be real below the $\pi\pi$ and $\pi\gamma$
thresholds.  Note that, from Eqs.~(\ref{five}) and (\ref{six}), any deviation
from the Breit-Wigner form and/or any non-linearity in the $q^2$-dependence of
$\Pi_{\rho\omega}(q^2)$ will produce a non-zero off-diagonal element of the
vector propagator {\it even in the physical basis}.  This means that a
background (non-resonant) term is completely unavoidable even in the
traditional VMD framework, where all contributions are associated with
vector meson exchange.  Moreover, in general, this background will be $s$-
(i.e., $q^2$)-dependent.  Finally, even in the vicinity of the $\rho$ and
$\omega$ poles, where it should be 
reasonable to set $\Pi_{\rho \rho}^{(0)} (q^2)$
and $\Pi_{\omega \omega}^{(0)} (q^2)$ 
to zero, the $\rho_I$ admixture into the
physical $\omega$ is governed, not by $\Pi^{\rho\omega} (m^2_\omega)$ as
usually assumed, but by $\Pi^{\rho\omega}(m^2_\rho)$.


The time-like EM pion form-factor is given, 
in the interference region, by
\be
{F}_\pi(q^2)=\left[g_{\omega\pi\pi}D_{\omega\omega}
\frac{f_{\omega\gamma}}{e}+
g_{\rho\pi\pi}D_{\rho\rho}\frac{f_{\rho\gamma}}{e}+g_{\rho\pi\pi}
D_{\rho\omega}\frac{f_{\omega\gamma}}{e}\right] + {\rm background},
\label{amp}
\ee
where $g_{\omega\pi\pi}$ is the coupling of the {\it physical} omega to the two
pion final state and $f_{\rho\gamma}$ and $f_{\omega\gamma}$ are the
electromagnetic $\rho$ and $\omega$ couplings.  The third piece of
Eq.~(\ref{amp}), $g_{\rho\pi\pi} D_{\rho\omega} f_{\omega\gamma}$, results from
the non-vanishing of the off-diagonal element of the {\em physical} meson
propagator and, being non-resonant, can be absorbed into the background, for
the purposes of our discussion, as can any deviations from the Breit-Wigner
form for the $\rho$ and $\omega$ propagators.  Since the variation of $q^2$
over the interference region is tiny, we can presumably also safely neglect any
$q^2$-dependence of $f_{\rho\gamma}$, $f_{\omega\gamma}$, $g_{\rho\pi\pi}$ and
$g_{\omega\pi\pi}$.  
$f_{V\gamma}$ is related to the ``universality
coupling''\cite{OWBK}, $g_V$, of traditional VMD treatments
by $f_{V\gamma}=-e\hat{m}^2/g_V$.  

	We now focus on the resonant $\omega$ exchange
contribution, whose magnitude and phase, relative
to the resonant $\rho$ exchange, are extracted
experimentally.  We have
\be
g_{\omega\pi\pi}=\langle\pi\pi|\omega_I+\epsilon_2\rho_I\rangle 
=g_{\omega_I\pi\pi}+\epsilon_2g_{\rho_I\pi\pi},
\label{omega}
\ee
where $\epsilon_2$ is given in Eq.~(\ref{six}) or, 
equivalently, by 
$\epsilon_2=-i\, z\Pi_{\rho\omega}(m_\rho^2)/
{\hat{m}_\rho\Gamma_\rho}$,
where
\be
z\equiv\left[1-\frac{\hat{m}_\omega\Gamma_\omega}{\hat{m}_\rho\Gamma_\rho}
-i\left(\frac{\hat{m}_\omega^2-\hat{m}_\rho^2}{\hat{m}_\rho\Gamma_\rho}\right)
\right]^{-1}.
\ee
Note that $z\approx 1$ but equals $1$ only if we neglect 
the $\omega$
width and $\rho-\omega$ mass difference.
This brings us to the Renard argument \cite{Renard}. 
Since, in general, $g_{\omega_I\pi\pi}\neq 0$,
$\Pi_{\rho\omega}(q^2)$ must contain a contribution from 
the intermediate $\pi\pi$ state which, because
essentially the entire $\rho$ width is due to the $\pi\pi$
mode, is given by
\be
\Pi^{2\pi}_{\rho\omega}(m_\rho^2)=
\frac{g_{\omega_I\pi\pi}}{g_{\rho_I\pi\pi}}\Pi^{2\pi}_{\rho\rho}(m_\rho^2)=G(
{\rm Re}\Pi^{2\pi}_{\rho\rho}(m_\rho^2)-i\hat{m}_\rho\Gamma_\rho),
\label{wcoupling}
\ee
where $G={g_{\omega_I\pi\pi}}/{g_{\rho_I\pi\pi}}$ is the ratio of 
the $\rho_I$ and $\omega_I$ couplings to $\pi\pi$. In arriving
at Eq.~\ref{wcoupling} we have
used the facts that (1) the imaginary part of the $\rho$ 
self-energy at resonance ($q^2=m_\rho^2$)
is, by definition, $-\hat{m}_\rho\Gamma_\rho$, and (2) 
$g_{\rho\pi\pi}=g_{\rho_I\pi\pi}$ to \oeps .
We have then, defining $\tilde\Pi_{\rho\omega}$ by
$\Pi_{\rho\omega}=\tilde\Pi_{\rho\omega}-iG{\hat m}\Gamma_\rho$,
\be
\epsilon_2=z\frac{-i}{\hat{m}_\rho\Gamma_\rho}[\tilde{\Pi}_{\rho\omega}(
m^2_\rho)-iG
\hat{m}_{\rho}\Gamma_\rho]
\label{neweps2}
\ee
and hence
\be
g_{\omega\pi\pi}=g_{\omega_I\pi\pi}\left(1-z\right)+
\tilde{\epsilon}_2g_{\rho_I\pi\pi},
\label{physomega}
\ee
where $\tilde{\epsilon}_2 =(-iz/ \hat{m}_\rho \Gamma_\rho) \tilde{\Pi}_{\rho
\omega} (m_\rho^2)$.
We shall also define, for convenience,
\be
\tilde{T}\equiv\tilde{\Pi}_{\rho\omega}
(m_\rho^2)/\hat{m}_\rho\Gamma_\rho.\ee
The standard Renard analysis \cite{Renard}
involves approximating $z$ by $1$.
The contribution to $\omega\ra\pi\pi$ from the intrinsic 
$\omega_I$ decay is then exactly cancelled
in Eq.~(\ref{physomega}).  Using the (preferred) experimental 
analysis of Ref.~\cite{Bernicha}, however, we find
\be
z=0.9324+0.3511\:i\ . \label{bern} 
\ee
(For comparison, the analysis of
Ref.~\cite{Benayoun} gives $1.023+0.2038\:i$).
Because of the substantial imaginary 
part,
the intrinsic decay cannot be neglected 
in \ep.

Substituting the results above into Eq.~(\ref{amp}), we find 
\be
{F}_\pi(q^2)=\frac{f_{\rho\gamma}}{e} g_{\rho_I\pi\pi}
\left[|r_{\rm ex}|e^{i\phi_{e^+e^-}}
\left((1-z)G-{\rm i}z\tilde{T}\right)P_\omega
+P_\rho\right] +\;{\rm background},
\label{complete}
\ee
where we have replaced the propagators 
$D_{\rho\rho ,\omega\omega}$ of
Eq.~(\ref{amp}) with the simple Breit-Wigner pole terms 
$P_{\rho ,\omega}\equiv 1/(p^2-m^2_{\rho ,\omega })$,
and where
\be
r_{\rm ex}\equiv\frac{f_{\omega\gamma} }{f_{\rho\gamma}}
=|r_{\rm ex}|e^{{\rm i}\phi_{e^+e^-}},
\label{rex1}
\ee
with $\phi_{e^+e^-}$ the ``leptonic phase'' (to be
discussed in more detail below).  
Experimentally,
\be
|r_{\rm ex}|=\left[\frac{\hat{m}^3_\omega\Gamma(\omega\ra e^+e^-)}
{\hat{m}^3_\rho
\Gamma(\rho\ra e^+e^-)}\right]^{1/2}
=0.30\pm 0.01
\ee
using the values found in Ref.~\cite{Bernicha}.
The form of $F_\pi(q^2)$ in Eq.~(\ref{complete}) is what is required for
comparison with experimental data \cite{Bernicha}, for
which one has
\be
F_\pi\propto P_\rho +Ae^{i\phi}P_\omega;\hskip0.5cm
A=-0.0109\pm 0.0011 ;\hskip0.5cm
\phi=(116.7\pm 5.8)^{\rm o}.
\label{fit}
\ee
One can now see that the uncertainty in the Orsay phase, 
$\phi$,
makes a precise extraction of $\tilde{\Pi}_{\rho\omega}(m_\rho^2)$ 
impossible.
Indeed, the two
contributions to the $\omega$ exchange amplitude 
(i.e., multiplying $P_\omega$) have either
nearly the same phase or differ in phase by close to $\pi$
(depending on the relative signs of $G$ and $\tilde T$).  In
either case, a large range of combinations of $G$ and $\tilde T$,
all producing nearly the same overall phase, will produce the
same value of $A$.
The experimental data can thus place only rather 
weak constraints on the
relative size of the two contributions, as we will see more
quantitatively below.

Let us write $r_{\rm ex}$, the ratio of electromagnetic couplings, 
in terms of the corresponding 
isospin-pure ratio, $r_I=f_{\omega_I\gamma}/
f_{\rho_I\gamma}$.
Using $f_{\omega\gamma}=f_{\omega_I\gamma} +
\epsilon_2 f_{\rho_I\gamma}$ and
$f_{\rho\gamma}= f_{\rho_I\gamma} -\epsilon_1 f_{\omega_I\gamma}$,
one finds
$r_{\rm ex}=(r_I+\epsilon_2)/(1-\epsilon_1 r_I)$,
where $r_I$ is real.  To \oeps\  one then has
\be
\sin\phi_{e^+e^-}=\frac{{\rm Im}(\epsilon_2)+|r_{\rm ex}|^2
{\rm Im}(\epsilon_1)}
{|r_{\rm ex}|}.
\label{philep}
\ee
Ignoring the small difference in $\epsilon_1$ and $\epsilon_2$
(since $r_{\rm ex}^2$ is small) we obtain
\be
\sin\phi_{e^+e^-} =\frac{(1+|r_{\rm ex}|^2) {\rm Im}\epsilon_2}{
|r_{\rm ex}|}.
\label{sinphi}
\ee

In order to simplify the discussion of our main point, which
is the effect of including the direct coupling on the 
experimental analysis, let us now make the
usual assumption that the imaginary part of $\Pi_{\rho\omega}$
is dominated by $\pi\pi$ intermediate states.  (Note, however, 
that, because the argument is complex, there may be an
imaginary part of $\Pi_{\rho\omega}$ even in the absence
of real intermediate states; for example, in the model of
Ref.\cite{MTRC}, with confined quark propagators, the phase
of the quark loop contribution to $\Pi_{\rho\omega}(m_\rho^2)$
is about $-13^{\rm o}$\cite{private}, 
despite the model having, for this contribution,
no available intermediate states.) 
Making
this assumption, $\tilde{\Pi}_{\rho\omega}$ (and thus $\tilde{T}$)
becomes pure real and the imaginary
part of $\Pi_{\rho\omega}(m_\rho^2)$ reduces to $-G\hat{m}_\rho
\Gamma_\rho$. Using Eqs.~(\ref{neweps2}) and (\ref{sinphi})
the leptonic phase becomes
\be
\sin\phi_{e^+e^-}=-\left(\frac{1+|r_{\rm ex}|^2}{|r_{\rm ex}|}\right)
(\tilde{T}\:{\rm Re}\:z+G\:{\rm Im}\:z)
\label{phifinal}
\ee
which is completely fixed by $G$ and 
$\tilde{\Pi}_{\rho\omega}$.  
For each possible
$\tilde{\Pi}_{\rho\omega}$, only one solution for $G$
both gives the correct experimental magnitude for the
$\omega$ exchange amplitude ($A$) and has a phase
lying in the second quadrant, as required by experiment. Knowing
$\tilde{\Pi}_{\rho\omega}$ and $G$, Eqn.~(\ref{phifinal}) allows
us to compute the total phase, $\phi$. 
Those pairs $(\tilde{\Pi}_{\rho\omega} , G)$ 
producing the experimentally allowed ($A$, $\phi$) 
constitute our full solution set.


The results of the above analysis are presented in Fig.~1, where we
have used as input the results of the analysis of
Ref.~\cite{Bernicha}, for the reasons explained above.  The spread in $G$
values reflects the experimental error in $A$.  We will supplement the
experimental constraints by imposing the theoretical prejudice $-0.05<G<0.05$. 
We see that, barring theoretical input on the 
precise size of $G$, experimental data is
incapable of providing even reasonably precise constraints on the individual
magnitudes of $G$ and $\tilde{\Pi}_{\rho \omega} (m_\rho^2)$.  The reason for
this situation has been explained above.  If we fix $A$
at its central value, the experimental phase alone would
restrict $\tilde{\Pi}_{\rho \omega}(m_\rho^2)$ 
to the 
range $(-1090\ {\rm MeV}^2, -5980\ {\rm MeV}^2)$,
the $G$ constraint to the range $(-2290\ {\rm MeV}^2, -6180\ {\rm MeV}^2)$.  
Including the experimental error on $A$ extends, for example, 
the phase constraint range to
$(-840\ {\rm MeV}^2, -6240\ {\rm MeV}^2)$.
For comparison, artificially setting
$G=0$ produces 
$\tilde{\Pi}_{\rho \omega}(m_\rho^2)=
-3960\ {\rm MeV}^2$.  
One may repeat the above analysis using the input 
parameters of
Ref.~\cite{Benayoun} (where, however, the $\rho$ 
pole position 
is presumably high by about $10\ {\rm MeV}$\cite{Bernicha}).
For the central $A$ value, the experimentally allowed range of
$\tilde{\Pi}_{\rho\omega}(m_\rho^2)$ is 
$(-3720\ {\rm MeV}^2, -5080\ {\rm
MeV}^2)$.  The large uncertainty in the extracted values of
$\tilde{\Pi}_{\rho\omega}(m_\rho^2)$ and $G$ is thus 
not an artifact of the
particular fit of Ref.~\cite{Bernicha}.  The small ($\pm 600\ {\rm MeV}^2$)
error usually quoted for $\tilde{\Pi}_{\rho\omega}(m_\rho^2)$, and associated
with the experimental error in the determination of $A$, thus represents a
highly inaccurate statement of the true uncertainty in the extraction of this
quantity from the experimental data.  It is important to stress that no further
information on $\tilde{\Pi}_{\rho\omega}(m_\rho^2)$ is obtainable from the \ep
data without additional theoretical input.  

Note that, in the model of Ref.~\cite{MTRC}, as currently parametrized, the
sign of $G$ is determined to be positive, and the magnitude to be $\simeq
0.02$.  Such a value of $G$, however, coupled with the phase correction
mentioned above, would fail to satisfy the experimental phase constraint.  This
shows that, despite the weakness of the experimental constraints for the
magnitudes of $G$ and $\tilde{\Pi}_{\rho\omega}(m_\rho^2)$, the experimental
results are, nonetheless, still capable of providing non-trivial constraints
for models of the mixing.


In conclusion, we have shown that, in general, 
there is a contribution to the 
$\rho\! -\!\omega$
interference in $e^+e^-\ra \pi^+\pi^-$ which arises
from the intrinsic $\omega_I \ra \pi\pi$ coupling, and that this contribution,
given the current level of accuracy of the experimentally extracted Orsay
phase, precludes any even reasonably precise extraction of the \rw mixing in
the absence of additional theoretical input.  
It is important to stress that this conclusion and the central result of
Eq.~(\ref{complete}) do not depend in the
least on the possible $q^2$-dependence of $\Pi_{\rho\omega}(q^2)$ nor on
the use of the $S$-matrix formalism:  even for constant $\Pi_{\rho\omega}$
and a more traditional Breit-Wigner analysis
one would still have a significant imaginary part of $z$ and hence
a residual contribution from the direct coupling which, being
nearly parallel to that associated with $\rho\! -\!\omega$ mixing,
would lead also to the conclusion stated above.
Note, however, that a significant
improvement in the determination of the experimental phase would allow one to
simultaneously extract the self-energy and the isospin-breaking ratio, $G$.  In
addition to the main point, just discussed, we also note that
(1) even if $G$ were, for some reason, to be zero, 
the data would provide the value of the 
mixing amplitude at $m^2_\rho$ and not $m^2_\omega$,
(2) since it is the complex S-matrix pole positions of
the $\rho$ and $\omega$ which govern the mixing parameters $\epsilon_{1,2}$,
only an analysis utilizing the S-matrix formalism can provide reliable input
for these pole positions, and hence for the analysis of the isospin-breaking
interference in \ep and (3) the simultaneous use of the experimental
magnitude and phase can provide non-trivial constraints on models of the vector
meson mixing process.  

\vspace{.5cm}

{\bf Acknowledgements:} We would like to thank M. Benayoun for discussions on
the experimental situation and K. Mitchell and P. Tandy
for providing information about the phase of the
\rw mixing amplitude off the real axis in the model
of Ref.~\cite{MTRC} prior to
publication. We also acknowledge helpful discussions
with A. Thomas.

\newpage

\begin{figure}[htb]
  \centering{\
     \epsfig{angle=0,figure=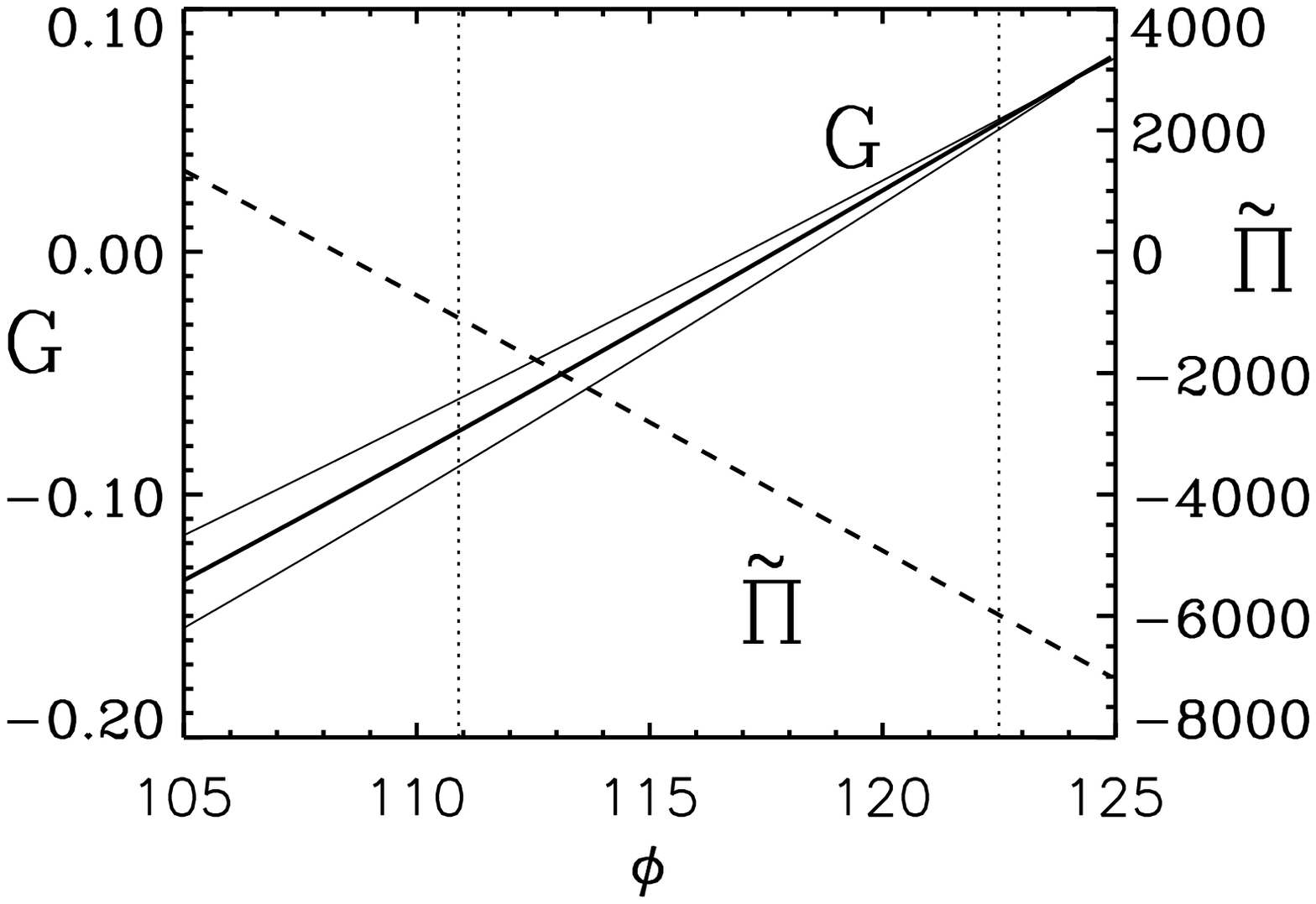,height=8.5cm}
               }
\parbox{130mm}{\caption
{The allowed values of \protect{$G=g_{\omega_I\pi\pi}/g_{\rho_I\pi\pi}$}
and \protect{$\tilde{\Pi}(m_\rho^2)$} (in MeV$^2$) are plotted as 
a function of the Orsay phase, $\phi$. The vertical lines indicate
the experimental uncertainty in $\phi$ 
\protect{$(=116.7\pm5.8)^{\rm o}$} and the uncertainty in the amplitude
$A$ ($0.0109\pm0.0011$) (see text) gives rise to the spread of possible
values of $G$ at each value of $\phi$.
}
\label{graph}}
\end{figure}

\end{document}